\documentclass[a4paper,12pt]{article}
\usepackage{mathptmx}
\usepackage{epsfig}
\usepackage{float}
\usepackage{amssymb}
\usepackage{latexsym}
\usepackage{authblk}

\usepackage{amsfonts}
\newcommand{\beq}{\begin{equation}}
\newcommand{\eeq}{\end{equation}}
\newcommand{\beqa}{\begin{eqnarray}}
\newcommand{\eeqa}{\end{eqnarray}}

\usepackage{natbib}
\usepackage{graphicx}
\usepackage{multirow}
\begin{document}



\title{Atmospheric ionization induced by precipitating electrons: Comparison of CRAC:EPII model with parametrization model}
\author[1]{A.A. Artamonov} 
\author[2]{A.L. Mishev}
\author[1,2]{I.G. Usoskin}
\affil[1]{Sodankyl\"a Geophysical Observatory (Oulu unit), University of Oulu, Finland.}
\affil[2]{ReSolve Center of Excellence, University of Oulu, Finland.}
\maketitle
\begin{abstract}

A new model CRAC:EPII (Cosmic Ray Atmospheric Cascade: Electron Precipitation Induced Ionization) is presented.  The CRAC:EPII is based on Monte Carlo simulation of precipitating electrons propagation and interaction with matter in the Earth atmosphere. It explicitly considers energy deposit: ionization, pair production, Compton scattering, generation of Bremsstrahlung high energy photons, photo-ionization and annihilation of positrons, multiple scattering as physical processes accordingly. 
The propagation of precipitating electrons and their interactions with atmospheric molecules is carried out with the GEANT4 simulation tool PLANETOCOSMICS code using NRLMSISE 00 atmospheric model. The ionization yields is compared with an analytical parametrization for various energies of incident precipitating electron, using a flux of mono-energetic particles. A good agreement between the two models is achieved. Subsequently, on the basis of balloon-born measured spectra of precipitating electrons at 30.10.2002 and 07.01.2004, the ion production rate in the middle and upper atmosphere is estimated using the CRAC:EPII model

\end{abstract}

\small Keywords:Atmospheric ionization, Stratosphere and Troposphere, Precipitation Electrons, Monte Carlo
solution
 \normalsize

\label{cor}{\small For contact: anton.artamonov@oulu.fi; alexander.mishev@oulu.fi; ilya.usoskin@oulu.fi}


\section{Introduction}

The main source of ionization in the troposphere and stratosphere is due to cosmic rays (CRs), which induce a complicated nuclear-electromagnetic-muon cascade resulting in an ionization of the ambient air \citep{Bri70, Usoskin2006, Bazilevskaya2008149, Stozhkov20091124, Velinov2013, Mironova2015}. Most of CR are protons and $\alpha-$ particles originating from outer space \citep{Nak10}. Small amounts of heavier nuclei are also present. Their intensity is modulated by the solar wind and heliomagnetic field, and follows the 11-year solar cycle. In addition, their flux responds to transient phenomena such as Forbush decreases \citep{Forbush37, Forbush58}.

In addition to energetic CR particles, a softer electron component of corpuscular radiation is present in the near-Earth space, which also ionizes the atmosphere, and specifically its upper part (see \cite{Li2001569, Millan2007362, Mironova2015} and references therein). Precipitation of electrons into the atmosphere occurs from various regions of the magnetosphere resulting from different mechanisms, some of them still poorly understood \citep[e.g.][and references therein]{Dorman04, Mironova2015}. Precipitating electrons play an important role in ion production in the Earth's atmosphere, specifically in the upper atmosphere over polar regions \citep{Makhmutov20031087, Daae2012, Clilverd20136921}. The precipitating electrons affect the atmospheric chemistry  \citep[e.g.][]{Rozanov20051,Verronen2011,Daae2012,Mironova2015} as well as several physical properties of the atmosphere and magnetosphere \citep[e.g.][]{Makhmutov20031087, Clilverd2008, JGRD:JGRD50518}. The intensity of the electron precipitation depends on solar activity \citep{Neal20152194, Makhmutov2001403}, season \citep{Makhmutov20031449}, geomagnetic activity \citep{Park2013247, Rodger2007, Horne2009} and other factors \citep{Makhmutov2006990}. Therefore, convenient model for assessment of atmospheric ionization due to  precipitating electrons as well as observations of energetic particles will stimulate better understanding of the impact of energetic particle to atmospheric processes \citep[e.g.][and references therein]{Mironova2015}.

Interactions between precipitating electrons and the atmosphere can be either parametrized using an analytical solution \citep[e.g.][]{Fang2008, Fang2010, McGranaghan2015} or can be modelled by a Monte Carlo method similarly to \citet{Solomon1993, Wissing2009, Wissing2011}. The parametrization models do not consider Bremsstrahlung, which contribute to ionization of air specifically at lower and middle altitudes, and consider direct ionization neglecting secondary process \citep[e.g.][]{Fang2008}. On the other hand, Monte Carlo transport codes consider realistically all the physics processes involved. In addition, models based on response (yield) function formalism, using precomputed ionization yields are more flexible compared to direct simulation, since not weighting of high energy particles simulation is necessary \citep[e.g.][]{Usoskin2006, Velinov2013, Mishev14d}. 

In this work, we present a new model for assessment of atmospheric ionization due to precipitating electrons and compare it with a previously proposed parametrization model. The general aim of this work is a quantitative comparison and demonstration of the ability of the new model to estimate the electron impact ionization. The new model, whose detailed description is given elsewhere, is an extension of the CRAC model for CR induced ionization \citep{Usoskin2006}, based on the ionization yield function formalism. It is a full target model similar to CRAC model for cosmic ray induced ionization and other similar models based on a Monte Carlo simulation of the atmospheric cascade \citep[e.g.]{Desorgher20056802, Usoskin2006, Velinov2009}.

\section{The model CRAC:EPII}
As was stated above, Monte Carlo simulation possess an advantage compared to parametrization by considering realistically all the physics processes involved. Here we apply Monte Carlo simulation of electron propagation and interaction with matter in the Earth atmosphere. The main advantage of Monte Carlo transport codes is that they consider, in a realistic manner, the physics processes, namely energy deposit, ionization, pair production, Compton scattering, generation of Bremsstrahlung high energy photons, photo-ionization and annihilation of positrons. Moreover the multiple scattering of electrons is realistically considered. In addition, since electrons produce Bremsstrahlung photons which penetrate deeper in the atmosphere, compared to primary particles \citep[e.g.][]{Schroter2006}, and produce ionization there, it is important to use adequate modelling of their production and propagation.

In this work the propagation of precipitating electrons and their interactions with atmospheric molecules, leading to production of secondary particles, is modelled using the GEANT4 based  \citep{Agostinelli2003250} simulation tool PLANETOCOSMICS \citep{Desorgher20056802}. Here we use a realistic curved atmospheric model NRLMSISE 00 \citep{Picone2002}. The code represents a Monte Carlo simulation tool for detailed study of cascade evolution in the atmosphere initiated by various primary particles. The code simulates the interactions and, where appropriate, decays of nuclei, hadrons, muons, electrons and photons in the atmosphere up to high and very high energies. It gives detailed information about the secondary particle flux at a selected observation level and the energy deposit, explicitly considering particle attenuation. The PLANETOCOSMICS also allows simulation of a purely electromagnetic cascade in a realistic manner.

We have computed the ionization yields (response function) i.e. the number of ion pairs produced per gram of the ambient air at a given atmospheric depth
by a single primary precipitating electron with a given energy. The computations were carried out in the energy range between 50 keV and 500 MeV. An example of ionization yields for several energies of primary electron is given in Fig.1. 

The ionization yields $Y$ given as ion $\mathrm{pairs\cdot cm^2g^{-1}}$, which corresponds to the atmospheric depth $x$, is defined as:

\begin{equation}
Y(x,K)=\frac{\Delta E}{E_i \Delta x}
\label{simp_eqn1}
   \end{equation}
   
\noindent where $\Delta E$ is the mean energy loss in the atmospheric layer $\Delta x$ centred at the atmospheric depth $x$ per one simulated primary electron with the kinetic energy $K$, and $E_i$=35 eV is the average energy needed to produce one ion pair \citep{Porter1976154}.

The ionization yields $Y(x,K)$ is related to the ion production rate Q(x) at a given depth $x$ as:
\begin{equation}
Q(x)=\int_{E_i}^\infty \frac{dJ_{e}}{dK} Y(x,K)\rho(x) dK
\label{simp_eqn2}
   \end{equation}

\noindent where $\frac{dJ_{e}}{dK} $ is the differential energy spectrum of the primary precipitating electrons with energy $K$, $\rho(x)$ is the atmospheric density at given atmospheric depth $x$.
As expected the maximum of ionization yields strongly depends on the energy of the precipitating electron. The maximum is lower for electrons with greater energy (Fig.1). In addition, significant fluctuations, specifically in a low energy range, of ionization yields are observed in the upper atmosphere at altitudes of about 90 km a.s.l. They are most-likely due to cascade to cascade development and/or attenuation fluctuations, rather than insufficient number of test particles in
the model run.

\begin{figure}[H]
\begin{center}
\epsfig{file=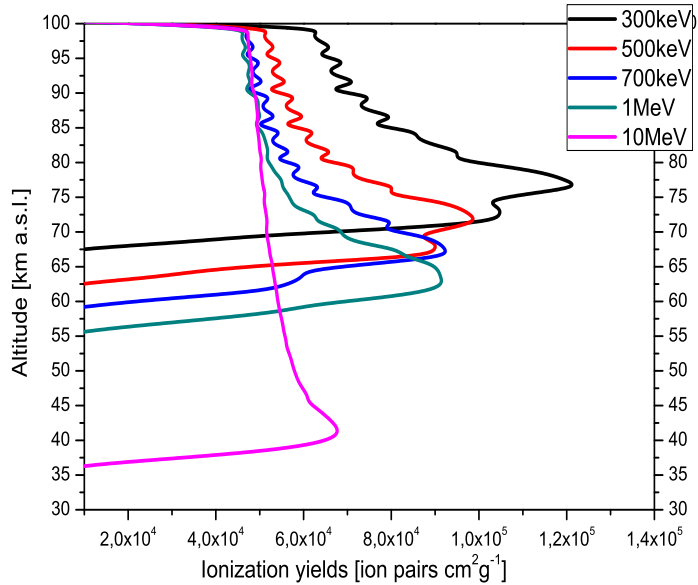,width=11cm, height=9cm} \caption{Ionization yields for primary electron with various energy as a function of the altitude above the sea level computed with CRAC:EPII model. The propagation of precipitating electrons in the atmosphere is modelled with the PLANETOCOSMICS code \citep{Desorgher20056802} using NRLMSISE 00 \citep{Picone2002}.}
\end{center}
\end{figure}

\section{Comparison with a parametrization model}
There are several parametrization models for assessment of electron induced ion production in the atmosphere \citep[e.g.][]{Lazarev1967, Roble1987, Frahm1997, Fang2010}.
Some of the models were focused on evaluation of auroral electron impact ionization \citep[e.g.][]{Roble1987}. In order to assess production of $NO_{x}$, in the middle and upper atmosphere, by high energy  electron precipitation \citep[e.g.][]{Callis19912939, Callis19961901, Aikin199926457, Turunen2009, Clilverd2010, Andersson2012, Krivolutsky2012685} an extension of parametrization models have been proposed \citep{Millan2007362,Fang2010}. Here we compare a mono-energetic ionization yields with a recent parametrization by \citep{Fang2008, Fang2010}, recently used for computation of ionization due to particle precipitation \citep{Huang2014}.

We compare the ionization yields due to high energy precipitating electrons (monoenergetic electron fluxes of 1 erg cm$^{-2}$ s$^{-1}$ propagating in the atmosphere according to \citet{Fang2010}), namely 100 keV and 1 MeV (Fig.2). The CRAC:EPII model predicts slightly more ions, specifically at the depth of maximum ion production. The observed difference in the maximum is of the order of 35-40 $\%$. The level of maximal ion production by CRAC:EPII is at slightly lower altitudes compared to parametrization model. In addition, the contribution of Bremsstrahlung photons to ionization is clearly seen at altitudes of about 30 km above the sea level. We achieve a satisfactory agreement with parametrization in the integral energy deposit. 

\begin{figure}[H]
\begin{center}
\epsfig{file=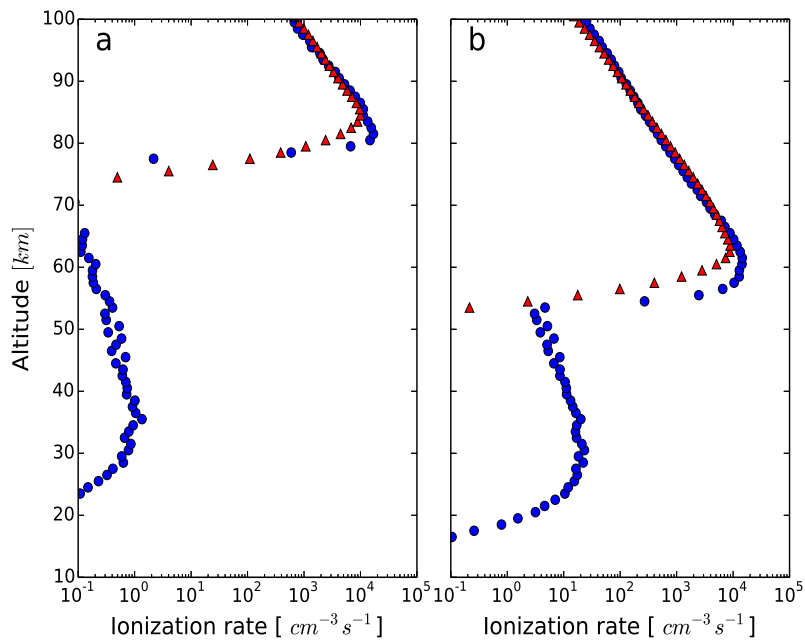,width=11cm, height=9cm} \caption{Comparison of ionization yields for primary electron with various energies as a function of the altitude above the sea level computed with CRAC:EPII model and parametrization model according to \citep{Fang2010}. Blue solid circles denote computations with CRAC:EPII, while red solid triangles the parametrization. a) electron flux of 1 erg cm$^{-2}$ s$^{-1}$ with energy of 100 keV; b) electron flux of 1 erg cm$^{-2}$ s$^{-1}$ with energy of 1 MeV.}
\end{center}
\end{figure}

The observed difference in the region of maximum ion production is due to a combination of various processes related to the complex high-energy electron propagation confined by the Monte Carlo model as well as the different atmospheric model assumptions. In general a good agreement between the two models is achieved, specifically in the upper atmosphere. Therefore, the CRAC:EPII demonstrate very good ability to assess ion production by high energy  electron precipitation. The CRAC:EPII accounts, in contrast to parametrization models, the contribution of Bremsstrahlung in the lower atmosphere, which is a important improvement.

\section{Spectrum of precipitating electrons and derived ion production rate}
At present various methods are proposed to estimate the spectrum of precipitating electrons \citep[e.g.][]{Clilverd2010, Neal20152194,Whittaker20137611, Wild2010}. In general, it is possible to reconstruct the spectra from satellite-born measurements \citep[e.g.][]{Rodger2010, Rodger2007, Peck20154596}. However, this method requires a correction as was recently shown by \citep{JGRA:JGRA50584}. Another possibility is proposed by \cite{Wild2010}. The detailed description of the precipitating electron spectra is beyond the topic of this work. Here we use the electron spectra obtained from balloon-born measurements \citep{Bazilevskaya19991670}, whose details are given in this volume \citep{Makhmutov2015}. In general it is assumed that the flux of precipitating electrons at the top of the atmosphere is exponential:

\begin{equation}
J_{e}(K)=A_{e}\cdot exp(-K/E_{0})
\label{simp_eqn3}
   \end{equation}

\citep[e.g.][]{Millan2007, Comess20135050}. The characteristic energy $E_{0}$ is in the range 10 keV - 1 MeV. The spectrum is reconstructed considering the characteristics of energetic electron precipitation in the polar atmosphere according to \citep{Makhmutov20031087}. Here we give an example of the event spectrum derived on the basis of balloon measurement in Murmansk region ($67^\circ33'N, 33^\circ20'E$) at 30.10.2002 and 07.01.2004. The characteristics of the spectra in Eq. 3 are $A_{e}$ = 9.37$\cdot$ 10$^{1}$ cm$^{-2}$ s$^{-2}$ keV$^{-1}$, $E_0$ = 3.09$\cdot$10$^{2}$ keV and $A_{e}$ = 4.96$\cdot$ 10$^{-1}$ cm$^{-2}$ s$^{-2}$ keV$^{-1}$, $E_0$ = 5.45$\cdot$10$^{3}$ keV, respectively, as shown in Fig.3. This spectra were used as input in the CRAC:EPII model in order to estimate the ion production rate (Fig.4). The derived ion production rate is in a good agreement with previous studies \citep[e.g.]{Makhmutov20031087, Sloan201129}.
It is clearly seen the contribution of Bremsstrahlung at altitudes below about 25 km a.s.l. for spectrum 1 and below 40 km a.s.l. for spectrum 2. 

\begin{figure}[H]
\begin{center}
\epsfig{file=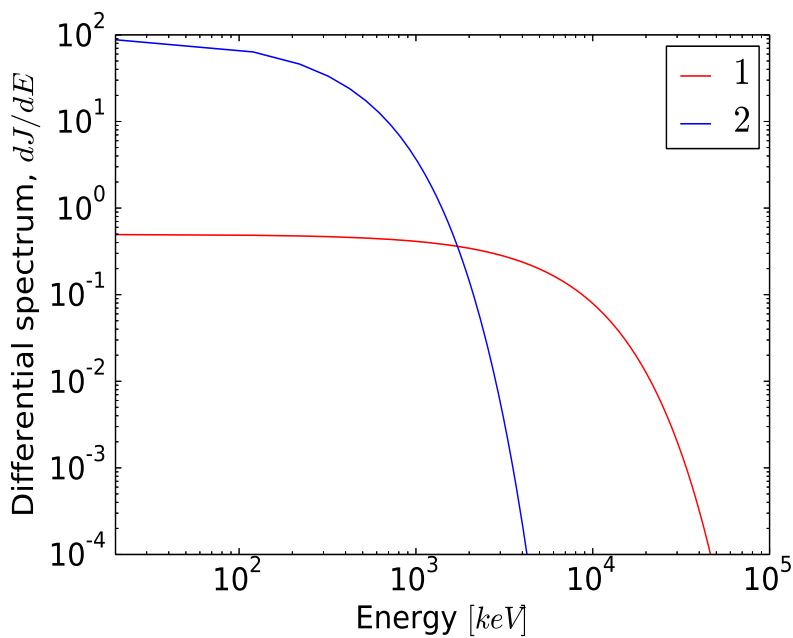,width=11cm, height=9cm} \caption{Example of the differential spectra of precipitating electrons. The spectra are derived on the basis of balloon measurement in Murmansk region ($67^\circ33'N, 33^\circ20'E$). The spectra characteristics are: $A_{e}$ = 9.37$\cdot$ 10$^{1}$ cm$^{-2}$ s$^{-2}$ keV$^{-1}$, $E_0$ = 3.09$\cdot$10$^{2}$ keV (at 30.10.2002, blue curve) and $A_{e}$ = 4.96$\cdot$ 10$^{-1}$ cm$^{-2}$ s$^{-2}$ keV$^{-1}$, $E_0$ = 5.45$\cdot$10$^{3}$ keV (07.01.2004, red curve).}
\end{center}
\end{figure}

\begin{figure}[H]
\begin{center}
\epsfig{file=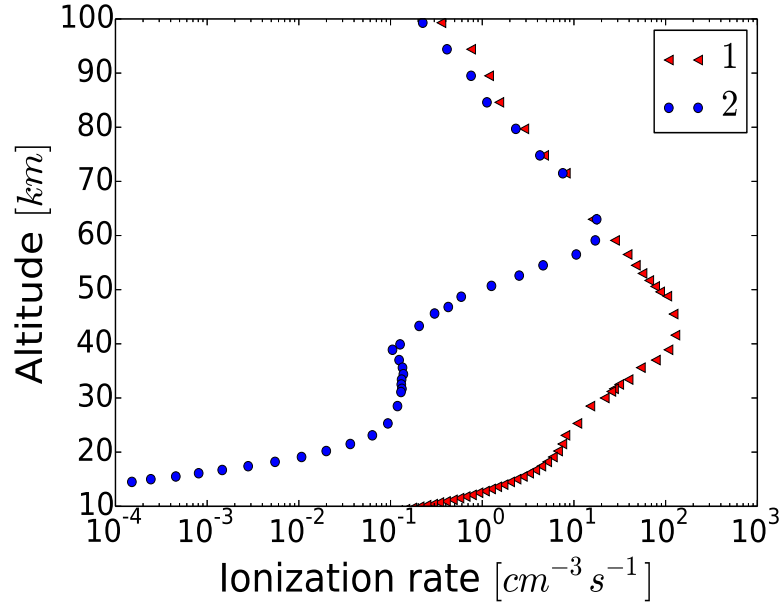,width=11cm, height=9cm} \caption{Ion production rate assuming the example spectra (Fig.3) of precipitating electrons (07.01.2004, red triangles) and (30.10.2002, blue dots) computed with CRAC:EPII model.}
\end{center}
\end{figure}

\section{Conclusion}
In this work we present a new model for assessment of atmospheric ionization induced by precipitating electrons and demonstrate a quantitative comparison with a parametrization model. The model is based on response (ionization yield) functions, derived with extensive Monte Carlo simulations. In contrast to parametrization models it accounts explicitly the contribution of Bremsstrahlung, important in the lower atmosphere. Moreover, it extend the energy range above 1 MeV (up to 500 MeV), which is the maximum energy in parametrization of \citep{Fang2008, Fang2010}.  In addition, compared to direct simulation models, it is more flexible (specifically for operational purposes), because it is based on a widely used, simple for application formalism of precomputed ionization yields.

We note that the present results show good agreement with a Monte Carlo model based on satellite measurements, AIMOS \citep{Wissing2009, Wissing2011}, which is based on GEANT4 \citep{Agostinelli2003250} (private communication with J. Wissing during the ISSI workshop on "Specification of ionization sources affecting atmospheric processes"). The direct comparison of these two models based on the same platform would be in practice a comparison of different atmospheric profiles resulting in particle transportation and/or different code versions,
but not comparison of different models. A detailed study of various atmospheric profile parametrizations and/or hadron generators within a same platform (tool) is discussed elsewhere \citep{Mishev14d} and it is beyond the scope of this work. 

The application of CRAC:EPII model for estimation of ionization yields demonstrates good agreement with a recent analytical parametrization model. In addition, CRAC:EPII was applied for ion rate production in the upper atmosphere using a balloon-born measured spectrum of precipitating electrons. A complete description of CRAC:EPII with the corresponding look-up tables of ionization yields, accordingly ionization yield function at several altitudes is presented elsewhere. 

\section*{Acknowledgements}
We warmly acknowledge the ISSI in Bern as well as all the colleagues from the ISSI project "Specification of ionization sources affecting atmospheric processes" lead by Irina Mironova. We acknowledge J. Wissing for the discussions related to AIMOS and ion production in the atmosphere due to precipitating electrons. We acknowledge G.A. Bazilevskaya and V.S. Makhmutov for the valuable and important notes. This work is supported by the Center of Excellence ReSoLVE  (project No. 272157 by Academy of Finland), University of Oulu Finland. 


\begin{thebibliography}{58}
\expandafter\ifx\csname natexlab\endcsname\relax\def\natexlab#1{#1}\fi
\providecommand{\url}[1]{\texttt{#1}}
\providecommand{\href}[2]{#2}
\providecommand{\path}[1]{#1}
\providecommand{\DOIprefix}{doi:}
\providecommand{\ArXivprefix}{arXiv:}
\providecommand{\URLprefix}{URL: }
\providecommand{\Pubmedprefix}{pmid:}
\providecommand{\doi}[1]{\href{http://dx.doi.org/#1}{\path{#1}}}
\providecommand{\Pubmed}[1]{\href{pmid:#1}{\path{#1}}}
\providecommand{\bibinfo}[2]{#2}
\ifx\xfnm\relax \def\xfnm[#1]{\unskip,\space#1}\fi
\bibitem[{Agostinelli et~al.(2003)Agostinelli, Allison, Amako, Apostolakis,
  Araujo, Arce, Asai, Axen, Banerjee, Barrand, Behner, Bellagamba, Boudreau,
  Broglia, Brunengo, Burkhardt, Chauvie, Chuma, Chytracek, Cooperman, Cosmo,
  Degtyarenko, Dell'Acqua, Depaola, Dietrich, Enami, Feliciello, Ferguson,
  Fesefeldt, Folger, Foppiano, Forti, Garelli, Giani, Giannitrapani, Gibin,
  Gomez~Cadenas, Gonzalez, Gracia~Abril, Greeniaus, Greiner, Grichine,
  Grossheim, Guatelli, Gumplinger, Hamatsu, Hashimoto, Hasui, Heikkinen,
  Howard, Ivanchenko, Johnson, Jones, Kallenbach, Kanaya, Kawabata, Kawabata,
  Kawaguti, Kelner, Kent, Kimura, Kodama, Kokoulin, Kossov, Kurashige, Lamanna,
  Lampen, Lara, Lefebure, Lei, Liendl, Lockman, Longo, Magni, Maire, Medernach,
  Minamimoto, Mora~de Freitas, Morita, Murakami, Nagamatu, Nartallo, Nieminen,
  Nishimura, Ohtsubo, Okamura, O'Neale, Oohata, Paech, Perl, Pfeiffer, Pia,
  Ranjard, Rybin, Sadilov, di~Salvo, Santin, Sasaki, Savvas, Sawada, Scherer,
  Sei, Sirotenko, Smith, Starkov, Stoecker, Sulkimo, Takahata, Tanaka,
  Tcherniaev, Safai~Tehrani, Tropeano, Truscott, Uno, Urban, Urban, Verderi,
  Walkden, Wander, Weber, Wellisch, Wenaus, Williams, Wright, Yamada, Yoshida
  and Zschiesche}]{Agostinelli2003250}
\bibinfo{author}{Agostinelli, S.}, \bibinfo{author}{Allison, J.},
  \bibinfo{author}{Amako, K.}, \bibinfo{author}{Apostolakis, J.},
  \bibinfo{author}{Araujo, H.}, \bibinfo{author}{Arce, P.},
  \bibinfo{author}{Asai, M.}, \bibinfo{author}{Axen, D.},
  \bibinfo{author}{Banerjee, S.}, \bibinfo{author}{Barrand, G.},
  \bibinfo{author}{Behner, F.}, \bibinfo{author}{Bellagamba, L.},
  \bibinfo{author}{Boudreau, J.}, \bibinfo{author}{Broglia, L.},
  \bibinfo{author}{Brunengo, A.}, \bibinfo{author}{Burkhardt, H.},
  \bibinfo{author}{Chauvie, S.}, \bibinfo{author}{Chuma, J.},
  \bibinfo{author}{Chytracek, R.}, \bibinfo{author}{Cooperman, G.},
  \bibinfo{author}{Cosmo, G.}, \bibinfo{author}{Degtyarenko, P.},
  \bibinfo{author}{Dell'Acqua, A.}, \bibinfo{author}{Depaola, G.},
  \bibinfo{author}{Dietrich, D.}, \bibinfo{author}{Enami, R.},
  \bibinfo{author}{Feliciello, A.}, \bibinfo{author}{Ferguson, C.},
  \bibinfo{author}{Fesefeldt, H.}, \bibinfo{author}{Folger, G.},
  \bibinfo{author}{Foppiano, F.}, \bibinfo{author}{Forti, A.},
  \bibinfo{author}{Garelli, S.}, \bibinfo{author}{Giani, S.},
  \bibinfo{author}{Giannitrapani, R.}, \bibinfo{author}{Gibin, D.},
  \bibinfo{author}{Gomez~Cadenas, J.}, \bibinfo{author}{Gonzalez, I.},
  \bibinfo{author}{Gracia~Abril, G.}, \bibinfo{author}{Greeniaus, G.},
  \bibinfo{author}{Greiner, W.}, \bibinfo{author}{Grichine, V.},
  \bibinfo{author}{Grossheim, A.}, \bibinfo{author}{Guatelli, S.},
  \bibinfo{author}{Gumplinger, P.}, \bibinfo{author}{Hamatsu, R.},
  \bibinfo{author}{Hashimoto, K.}, \bibinfo{author}{Hasui, H.},
  \bibinfo{author}{Heikkinen, A.}, \bibinfo{author}{Howard, A.},
  \bibinfo{author}{Ivanchenko, V.}, \bibinfo{author}{Johnson, A.},
  \bibinfo{author}{Jones, F.}, \bibinfo{author}{Kallenbach, J.},
  \bibinfo{author}{Kanaya, N.}, \bibinfo{author}{Kawabata, M.},
  \bibinfo{author}{Kawabata, Y.}, \bibinfo{author}{Kawaguti, M.},
  \bibinfo{author}{Kelner, S.}, \bibinfo{author}{Kent, P.},
  \bibinfo{author}{Kimura, A.}, \bibinfo{author}{Kodama, T.},
  \bibinfo{author}{Kokoulin, R.}, \bibinfo{author}{Kossov, M.},
  \bibinfo{author}{Kurashige, H.}, \bibinfo{author}{Lamanna, E.},
  \bibinfo{author}{Lampen, T.}, \bibinfo{author}{Lara, V.},
  \bibinfo{author}{Lefebure, V.}, \bibinfo{author}{Lei, F.},
  \bibinfo{author}{Liendl, M.}, \bibinfo{author}{Lockman, W.},
  \bibinfo{author}{Longo, F.}, \bibinfo{author}{Magni, S.},
  \bibinfo{author}{Maire, M.}, \bibinfo{author}{Medernach, E.},
  \bibinfo{author}{Minamimoto, K.}, \bibinfo{author}{Mora~de Freitas, P.},
  \bibinfo{author}{Morita, Y.}, \bibinfo{author}{Murakami, K.},
  \bibinfo{author}{Nagamatu, M.}, \bibinfo{author}{Nartallo, R.},
  \bibinfo{author}{Nieminen, P.}, \bibinfo{author}{Nishimura, T.},
  \bibinfo{author}{Ohtsubo, K.}, \bibinfo{author}{Okamura, M.},
  \bibinfo{author}{O'Neale, S.}, \bibinfo{author}{Oohata, Y.},
  \bibinfo{author}{Paech, K.}, \bibinfo{author}{Perl, J.},
  \bibinfo{author}{Pfeiffer, A.}, \bibinfo{author}{Pia, M.},
  \bibinfo{author}{Ranjard, F.}, \bibinfo{author}{Rybin, A.},
  \bibinfo{author}{Sadilov, S.}, \bibinfo{author}{di~Salvo, E.},
  \bibinfo{author}{Santin, G.}, \bibinfo{author}{Sasaki, T.},
  \bibinfo{author}{Savvas, N.}, \bibinfo{author}{Sawada, Y.},
  \bibinfo{author}{Scherer, S.}, \bibinfo{author}{Sei, S.},
  \bibinfo{author}{Sirotenko, V.}, \bibinfo{author}{Smith, D.},
  \bibinfo{author}{Starkov, N.}, \bibinfo{author}{Stoecker, H.},
  \bibinfo{author}{Sulkimo, J.}, \bibinfo{author}{Takahata, M.},
  \bibinfo{author}{Tanaka, S.}, \bibinfo{author}{Tcherniaev, E.},
  \bibinfo{author}{Safai~Tehrani, E.}, \bibinfo{author}{Tropeano, M.},
  \bibinfo{author}{Truscott, P.}, \bibinfo{author}{Uno, H.},
  \bibinfo{author}{Urban, L.}, \bibinfo{author}{Urban, P.},
  \bibinfo{author}{Verderi, M.}, \bibinfo{author}{Walkden, A.},
  \bibinfo{author}{Wander, W.}, \bibinfo{author}{Weber, H.},
  \bibinfo{author}{Wellisch, J.}, \bibinfo{author}{Wenaus, T.},
  \bibinfo{author}{Williams, D.}, \bibinfo{author}{Wright, D.},
  \bibinfo{author}{Yamada, T.}, \bibinfo{author}{Yoshida, H.},
  \bibinfo{author}{Zschiesche, D.}, \bibinfo{year}{2003}.
\newblock \bibinfo{title}{Geant4 - a simulation toolkit}.
\newblock \bibinfo{journal}{Nuclear Instruments and Methods in Physics
  Research, Section A: Accelerators, Spectrometers, Detectors and Associated
  Equipment} \bibinfo{volume}{506}, \bibinfo{pages}{250--303}.
\bibitem[{Aikin and Smith(1999)}]{Aikin199926457}
\bibinfo{author}{Aikin, A.}, \bibinfo{author}{Smith, H.}, \bibinfo{year}{1999}.
\newblock \bibinfo{title}{Mesospheric constituent variations during electron
  precipitation events}.
\newblock \bibinfo{journal}{Journal of Geophysical Research Atmospheres}
  \bibinfo{volume}{104}, \bibinfo{pages}{26457--26471}.
\bibitem[{Andersson et~al.(2012)Andersson, Verronen, Wang, Rodger, Clilverd and
  Carson}]{Andersson2012}
\bibinfo{author}{Andersson, M.}, \bibinfo{author}{Verronen, P.},
  \bibinfo{author}{Wang, S.}, \bibinfo{author}{Rodger, C.},
  \bibinfo{author}{Clilverd, M.}, \bibinfo{author}{Carson, B.},
  \bibinfo{year}{2012}.
\newblock \bibinfo{title}{Precipitating radiation belt electrons and
  enhancements of mesospheric hydroxyl during 2004-2009}.
\newblock \bibinfo{journal}{Journal of Geophysical Research Atmospheres}
  \bibinfo{volume}{117}, \bibinfo{pages}{D09304}.
\bibitem[{Asikainen and Mursula(2013)}]{JGRA:JGRA50584}
\bibinfo{author}{Asikainen, T.}, \bibinfo{author}{Mursula, K.},
  \bibinfo{year}{2013}.
\newblock \bibinfo{title}{Correcting the noaa/meped energetic electron fluxes
  for detector efficiency and proton contamination}.
\newblock \bibinfo{journal}{Journal of Geophysical Research: Space Physics}
  \bibinfo{volume}{118}, \bibinfo{pages}{6500--6510}.
\bibitem[{Bazilevskaya and Makhmutov(1999)}]{Bazilevskaya19991670}
\bibinfo{author}{Bazilevskaya, G.}, \bibinfo{author}{Makhmutov, V.},
  \bibinfo{year}{1999}.
\newblock \bibinfo{title}{Precipitations of energetic electrons into atmosphere
  according to the data using zond particle measurements}.
\newblock \bibinfo{journal}{Izvestiya Akademii Nauk. Ser. Fizicheskaya}
  \bibinfo{volume}{63}, \bibinfo{pages}{1670--1674}.
\bibitem[{Bazilevskaya et~al.(2008)Bazilevskaya, Usoskin, Fl{\"u}ckiger, Harrison,
  Desorgher, B{\"u}tikofer, Krainev, Makhmutov, Stozhkov, Svirzhevskaya,
  Svirzhevsky and Kovaltsov}]{Bazilevskaya2008149}
\bibinfo{author}{Bazilevskaya, G.}, \bibinfo{author}{Usoskin, I.},
  \bibinfo{author}{Fl{\"u}ckiger, E.}, \bibinfo{author}{Harrison, R.},
  \bibinfo{author}{Desorgher, L.}, \bibinfo{author}{B{\"u}tikofer, R.},
  \bibinfo{author}{Krainev, M.}, \bibinfo{author}{Makhmutov, V.},
  \bibinfo{author}{Stozhkov, Y.}, \bibinfo{author}{Svirzhevskaya, A.},
  \bibinfo{author}{Svirzhevsky, N.}, \bibinfo{author}{Kovaltsov, G.},
  \bibinfo{year}{2008}.
\newblock \bibinfo{title}{Cosmic ray induced ion production in the atmosphere}.
\newblock \bibinfo{journal}{Space Science Reviews} \bibinfo{volume}{137},
  \bibinfo{pages}{149--173}.
\bibitem[{Callis(1991)}]{Callis19912939}
\bibinfo{author}{Callis, L.}, \bibinfo{year}{1991}.
\newblock \bibinfo{title}{Precipitating relativistic electrons: their long-term
  effect on stratospheric odd nitrogen levels}.
\newblock \bibinfo{journal}{Journal of Geophysical Research}
  \bibinfo{volume}{96}, \bibinfo{pages}{2939--2976}.
\bibitem[{Callis et~al.(1996)Callis, Boughner, Baker, Mewaldt, Bernard~Blake,
  Selesnick, Cummings, Natarajan, Mason and Mazur}]{Callis19961901}
\bibinfo{author}{Callis, L.}, \bibinfo{author}{Boughner, R.},
  \bibinfo{author}{Baker, D.}, \bibinfo{author}{Mewaldt, R.},
  \bibinfo{author}{Bernard~Blake, J.}, \bibinfo{author}{Selesnick, R.},
  \bibinfo{author}{Cummings, J.}, \bibinfo{author}{Natarajan, M.},
  \bibinfo{author}{Mason, G.}, \bibinfo{author}{Mazur, J.},
  \bibinfo{year}{1996}.
\newblock \bibinfo{title}{Precipitating electrons: Evidence for effects on
  mesospheric odd nitrogen}.
\newblock \bibinfo{journal}{Geophysical Research Letters} \bibinfo{volume}{23},
  \bibinfo{pages}{1901--1904}.
\bibitem[{Clilverd et~al.(2013)Clilverd, Cobbett, Rodger, Brundell, Denton,
  Hartley, Rodriguez, Danskin, Raita and Spanswick}]{Clilverd20136921}
\bibinfo{author}{Clilverd, M.}, \bibinfo{author}{Cobbett, N.},
  \bibinfo{author}{Rodger, C.}, \bibinfo{author}{Brundell, J.},
  \bibinfo{author}{Denton, M.}, \bibinfo{author}{Hartley, D.},
  \bibinfo{author}{Rodriguez, J.}, \bibinfo{author}{Danskin, D.},
  \bibinfo{author}{Raita, T.}, \bibinfo{author}{Spanswick, E.},
  \bibinfo{year}{2013}.
\newblock \bibinfo{title}{Energetic electron precipitation characteristics
  observed from antarctica during a flux dropout event}.
\newblock \bibinfo{journal}{Journal of Geophysical Research: Space Physics}
  \bibinfo{volume}{118}, \bibinfo{pages}{6921--6935}.
\bibitem[{Clilverd et~al.(2008)Clilverd, Rodger, Brundell, B{\"a}hr, Cobbett,
  Moffat-Griffin, Kavanagh, Sepp{\"a}l{\"a}, Thomson, Friedel and
  Menk}]{Clilverd2008}
\bibinfo{author}{Clilverd, M.}, \bibinfo{author}{Rodger, C.},
  \bibinfo{author}{Brundell, J.}, \bibinfo{author}{B{\"a}hr, J.},
  \bibinfo{author}{Cobbett, N.}, \bibinfo{author}{Moffat-Griffin, T.},
  \bibinfo{author}{Kavanagh, A.}, \bibinfo{author}{Sepp{\"a}l{\"a}, A.},
  \bibinfo{author}{Thomson, N.}, \bibinfo{author}{Friedel, R.},
  \bibinfo{author}{Menk, F.}, \bibinfo{year}{2008}.
\newblock \bibinfo{title}{Energetic electron precipitation during substorm
  injection events: High-latitude fluxes and an unexpected midlatitude
  signature}.
\newblock \bibinfo{journal}{Journal of Geophysical Research: Space Physics}
  \bibinfo{volume}{113}, \bibinfo{pages}{A10311}.
\bibitem[{Clilverd et~al.(2010)Clilverd, Rodger, Gamble, Ulich, Raita,
  Sep{\"a}pl{\"a}, Green, Thomson, Sauvaud and Parrot}]{Clilverd2010}
\bibinfo{author}{Clilverd, M.}, \bibinfo{author}{Rodger, C.},
  \bibinfo{author}{Gamble, R.}, \bibinfo{author}{Ulich, T.},
  \bibinfo{author}{Raita, T.}, \bibinfo{author}{Sep{\"a}pl{\"a}, A.},
  \bibinfo{author}{Green, J.}, \bibinfo{author}{Thomson, N.},
  \bibinfo{author}{Sauvaud, J.A.}, \bibinfo{author}{Parrot, M.},
  \bibinfo{year}{2010}.
\newblock \bibinfo{title}{Ground-based estimates of outer radiation belt
  energetic electron precipitation fluxes into the atmosphere}.
\newblock \bibinfo{journal}{Journal of Geophysical Research: Space Physics}
  \bibinfo{volume}{115}, \bibinfo{pages}{A12304}.
\bibitem[{Comess et~al.(2013)Comess, Smith, Selesnick, Millan and
  Sample}]{Comess20135050}
\bibinfo{author}{Comess, M.}, \bibinfo{author}{Smith, D.},
  \bibinfo{author}{Selesnick, R.}, \bibinfo{author}{Millan, R.},
  \bibinfo{author}{Sample, J.}, \bibinfo{year}{2013}.
\newblock \bibinfo{title}{Duskside relativistic electron precipitation as
  measured by sampex: A statistical survey}.
\newblock \bibinfo{journal}{Journal of Geophysical Research: Space Physics}
  \bibinfo{volume}{118}, \bibinfo{pages}{5050--5058}.
\bibitem[{Daae et~al.(2012)Daae, Espy, Nesse~Tyssy, Newnham, Stadsnes and
  Sraas}]{Daae2012}
\bibinfo{author}{Daae, M.}, \bibinfo{author}{Espy, P.},
  \bibinfo{author}{Nesse~Tyssy, H.}, \bibinfo{author}{Newnham, D.},
  \bibinfo{author}{Stadsnes, J.}, \bibinfo{author}{Sraas, F.},
  \bibinfo{year}{2012}.
\newblock \bibinfo{title}{The effect of energetic electron precipitation on
  middle mesospheric night-time ozone during and after a moderate geomagnetic
  storm}.
\newblock \bibinfo{journal}{Geophysical Research Letters} \bibinfo{volume}{39},
  \bibinfo{pages}{L21811}.
\bibitem[{Desorgher et~al.(2005)Desorgher, Fl{\"u}ckiger, Gurtner, Moser and
  B{\"u}tikofer}]{Desorgher20056802}
\bibinfo{author}{Desorgher, L.}, \bibinfo{author}{Fl{\"u}ckiger, E.},
  \bibinfo{author}{Gurtner, M.}, \bibinfo{author}{Moser, M.},
  \bibinfo{author}{B{\"u}tikofer, R.}, \bibinfo{year}{2005}.
\newblock \bibinfo{title}{Atmocosmics: A geant 4 code for computing the
  interaction of cosmic rays with the earth's atmosphere}.
\newblock \bibinfo{journal}{International Journal of Modern Physics A}
  \bibinfo{volume}{20}, \bibinfo{pages}{6802--6804}.
\bibitem[{Dorman(2004)}]{Dorman04}
\bibinfo{author}{Dorman, L.}, \bibinfo{year}{2004}.
\newblock \bibinfo{title}{Cosmic Rays in the Earth's Atmosphere and
  Underground}.
\newblock \bibinfo{publisher}{Kluwer Academic Publishers},
  \bibinfo{address}{Dordrecht}.
\bibitem[{Fang et~al.(2008)Fang, Randall, Lummerzheim, Solomon, Mills, Marsh,
  Jackman, Wang and Lu}]{Fang2008}
\bibinfo{author}{Fang, X.}, \bibinfo{author}{Randall, C.},
  \bibinfo{author}{Lummerzheim, D.}, \bibinfo{author}{Solomon, S.},
  \bibinfo{author}{Mills, M.}, \bibinfo{author}{Marsh, D.},
  \bibinfo{author}{Jackman, C.}, \bibinfo{author}{Wang, W.},
  \bibinfo{author}{Lu, G.}, \bibinfo{year}{2008}.
\newblock \bibinfo{title}{Electron impact ionization: A new parameterization
  for 100 ev to 1 mev electrons}.
\newblock \bibinfo{journal}{Journal of Geophysical Research: Space Physics}
  \bibinfo{volume}{113}, \bibinfo{pages}{A09311}.
\bibitem[{Fang et~al.(2010)Fang, Randall, Lummerzheim, Wang, Lu, Solomon and
  Frahm}]{Fang2010}
\bibinfo{author}{Fang, X.}, \bibinfo{author}{Randall, C.},
  \bibinfo{author}{Lummerzheim, D.}, \bibinfo{author}{Wang, W.},
  \bibinfo{author}{Lu, G.}, \bibinfo{author}{Solomon, S.},
  \bibinfo{author}{Frahm, R.}, \bibinfo{year}{2010}.
\newblock \bibinfo{title}{Parameterization of monoenergetic electron impact
  ionization}.
\newblock \bibinfo{journal}{Geophysical Research Letters} \bibinfo{volume}{37},
  \bibinfo{pages}{L22106}.
\bibitem[{Forbush(1937)}]{Forbush37}
\bibinfo{author}{Forbush, S.}, \bibinfo{year}{1937}.
\newblock \bibinfo{title}{On the effects in cosmic-ray intensity observed
  during the recent magnetic storm}.
\newblock \bibinfo{journal}{Physical Review} \bibinfo{volume}{51},
  \bibinfo{pages}{1108}.
\bibitem[{Forbush(1958)}]{Forbush58}
\bibinfo{author}{Forbush, S.}, \bibinfo{year}{1958}.
\newblock \bibinfo{title}{Cosmic-ray intensity variations during two solar
  cycles}.
\newblock \bibinfo{journal}{Journal of Geophysical Research}
  \bibinfo{volume}{63}, \bibinfo{pages}{651--669}.
\bibitem[{Frahm et~al.(1997)Frahm, Winningham, Sharber, Link, Crowley, Gaines,
  Chenette, Anderson and Potemra}]{Frahm1997}
\bibinfo{author}{Frahm, R.}, \bibinfo{author}{Winningham, J.},
  \bibinfo{author}{Sharber, J.}, \bibinfo{author}{Link, R.},
  \bibinfo{author}{Crowley, G.}, \bibinfo{author}{Gaines, E.},
  \bibinfo{author}{Chenette, D.}, \bibinfo{author}{Anderson, B.},
  \bibinfo{author}{Potemra, T.}, \bibinfo{year}{1997}.
\newblock \bibinfo{title}{The diffuse aurora: A significant source of
  ionization in the middle atmosphere}.
\newblock \bibinfo{journal}{Journal of Geophysical Research Atmospheres}
  \bibinfo{volume}{102}, \bibinfo{pages}{28203--28214}.
\bibitem[{Gaisser and Stanev(2010)}]{Nak10}
\bibinfo{author}{Gaisser, T.K.}, \bibinfo{author}{Stanev, T.},
  \bibinfo{year}{2010}.
\newblock \bibinfo{title}{Cosmic rays}, in: \bibinfo{editor}{et~al., K.N.}
  (Ed.), \bibinfo{booktitle}{Review of Particle Physics}.
  \bibinfo{publisher}{Journal of Physics G 37}, pp. \bibinfo{pages}{269--275}.
\bibitem[{Horne et~al.(2009)Horne, Lam and Green}]{Horne2009}
\bibinfo{author}{Horne, R.}, \bibinfo{author}{Lam, M.}, \bibinfo{author}{Green,
  J.}, \bibinfo{year}{2009}.
\newblock \bibinfo{title}{Energetic electron precipitation from the outer
  radiation belt during geomagnetic storms}.
\newblock \bibinfo{journal}{Geophysical Research Letters} \bibinfo{volume}{36}.
\bibitem[{Huang et~al.(2014)Huang, Huang, Su, Deng and Fang}]{Huang2014}
\bibinfo{author}{Huang, Y.}, \bibinfo{author}{Huang, C.}, \bibinfo{author}{Su,
  Y.J.}, \bibinfo{author}{Deng, Y.}, \bibinfo{author}{Fang, X.},
  \bibinfo{year}{2014}.
\newblock \bibinfo{title}{Ionization due to electron and proton precipitation
  during the august 2011 storm}.
\newblock \bibinfo{journal}{Journal of Geophysical Research: Space Physics}
  \bibinfo{volume}{119}, \bibinfo{pages}{3106--3116}.
\bibitem[{Krivolutsky and Repnev(2012)}]{Krivolutsky2012685}
\bibinfo{author}{Krivolutsky, A.}, \bibinfo{author}{Repnev, A.},
  \bibinfo{year}{2012}.
\newblock \bibinfo{title}{Impact of space energetic particles on the earth's
  atmosphere (a review)}.
\newblock \bibinfo{journal}{Geomagnetism and Aeronomy} \bibinfo{volume}{52},
  \bibinfo{pages}{685--716}.
\bibitem[{Lazarev(1967)}]{Lazarev1967}
\bibinfo{author}{Lazarev, V.}, \bibinfo{year}{1967}.
\newblock \bibinfo{title}{Absorption of the energy of an electron beam in the
  upper atmosphere}.
\newblock \bibinfo{journal}{Geomagnetism and aeronomy} \bibinfo{volume}{7},
  \bibinfo{pages}{219--4949}.
\bibitem[{Li and Temerin(2001)}]{Li2001569}
\bibinfo{author}{Li, X.}, \bibinfo{author}{Temerin, M.}, \bibinfo{year}{2001}.
\newblock \bibinfo{title}{The electron radiation belt}.
\newblock \bibinfo{journal}{Space Science Reviews} \bibinfo{volume}{95},
  \bibinfo{pages}{569--580}.
\bibitem[{Makhmutov et~al.(2006)Makhmutov, Bazilevskaya, Desorgher and
  Fl{\"u}ckiger}]{Makhmutov2006990}
\bibinfo{author}{Makhmutov, V.}, \bibinfo{author}{Bazilevskaya, G.},
  \bibinfo{author}{Desorgher, L.}, \bibinfo{author}{Fl{\"u}ckiger, E.},
  \bibinfo{year}{2006}.
\newblock \bibinfo{title}{Observation of energetic electron precipitation into
  atmosphere in october 2003}.
\newblock \bibinfo{journal}{Bulletin of the Russian Academy of Sciences:
  Physics} \bibinfo{volume}{69}, \bibinfo{pages}{990--993}.
\bibitem[{Makhmutov et~al.(2003a)Makhmutov, Bazilevskaya and
  Krainev}]{Makhmutov20031087}
\bibinfo{author}{Makhmutov, V.}, \bibinfo{author}{Bazilevskaya, G.},
  \bibinfo{author}{Krainev, M.}, \bibinfo{year}{2003}a.
\newblock \bibinfo{title}{Characteristics of energetic electron precipitation
  into the earth's polar atmosphere and geomagnetic conditions}.
\newblock \bibinfo{journal}{Advances in Space Research} \bibinfo{volume}{31},
  \bibinfo{pages}{1087--1092}.
\bibitem[{Makhmutov et~al.(2001)Makhmutov, Bazilevskaya, Krainev, Svirzhevskaya
  and Svirzhevsky}]{Makhmutov2001403}
\bibinfo{author}{Makhmutov, V.}, \bibinfo{author}{Bazilevskaya, G.},
  \bibinfo{author}{Krainev, M.}, \bibinfo{author}{Svirzhevskaya, A.},
  \bibinfo{author}{Svirzhevsky, N.}, \bibinfo{year}{2001}.
\newblock \bibinfo{title}{Connection of frequency of precipitation of
  relativistic electrons to atmosphere with the solar activity cycle}.
\newblock \bibinfo{journal}{Izvestiya Akademii Nauk. Ser. Fizicheskaya}
  \bibinfo{volume}{65}, \bibinfo{pages}{403--405}.
\bibitem[{Makhmutov et~al.(2003b)Makhmutov, Bazilevskaya and
  Stozhkov}]{Makhmutov20031449}
\bibinfo{author}{Makhmutov, V.}, \bibinfo{author}{Bazilevskaya, G.},
  \bibinfo{author}{Stozhkov, Y.}, \bibinfo{year}{2003}b.
\newblock \bibinfo{title}{Seasonal effect in precipitation of energetic
  electrons into polar atmosphere}.
\newblock \bibinfo{journal}{Izvestiya Akademii Nauk. Ser. Fizicheskaya}
  \bibinfo{volume}{67}, \bibinfo{pages}{1449--1452}.
\bibitem[{Makhmutov et~al.(2015)Makhmutov, Bazilevskaya, Stozhkov,
  Svirzhevskaya and Svirzhevsky}]{Makhmutov2015}
\bibinfo{author}{Makhmutov, V.}, \bibinfo{author}{Bazilevskaya, G.},
  \bibinfo{author}{Stozhkov, Y.}, \bibinfo{author}{Svirzhevskaya, A.},
  \bibinfo{author}{Svirzhevsky, N.}, \bibinfo{year}{2015}.
\newblock \bibinfo{title}{Catalogue of electron precipitation events as
  observed in the long-duration cosmic ray balloon experiment}.
\newblock \bibinfo{journal}{Journal of Atmospheric and Solar-Terrestrial
  Physics} , \bibinfo{pages}{in press, doi:10.1016/j.jastp.2015.12.006}.
\bibitem[{Maliniemi et~al.(2013)Maliniemi, Asikainen, Mursula and
  Sepp{\"a}l{\"a}}]{JGRD:JGRD50518}
\bibinfo{author}{Maliniemi, V.}, \bibinfo{author}{Asikainen, T.},
  \bibinfo{author}{Mursula, K.}, \bibinfo{author}{Sepp{\"a}l{\"a}, A.},
  \bibinfo{year}{2013}.
\newblock \bibinfo{title}{Qbo-dependent relation between electron precipitation
  and wintertime surface temperature}.
\newblock \bibinfo{journal}{Journal of Geophysical Research: Atmospheres}
  \bibinfo{volume}{118}, \bibinfo{pages}{6302--6310}.
\bibitem[{McGranaghan et~al.(2015)McGranaghan, Knipp, Solomon and
  Fang}]{McGranaghan2015}
\bibinfo{author}{McGranaghan, R.}, \bibinfo{author}{Knipp, D.},
  \bibinfo{author}{Solomon, S.}, \bibinfo{author}{Fang, X.},
  \bibinfo{year}{2015}.
\newblock \bibinfo{title}{A fast, parameterized model of upper atmospheric
  ionization rates, chemistry, and conductivity}.
\newblock \bibinfo{journal}{Journal of Geophysical Research A: Space Physics}
  \bibinfo{volume}{120}, \bibinfo{pages}{4936--4949}.
\bibitem[{Millan et~al.(2007)Millan, Lin, Smith and McCarthy}]{Millan2007}
\bibinfo{author}{Millan, R.}, \bibinfo{author}{Lin, R.},
  \bibinfo{author}{Smith, D.}, \bibinfo{author}{McCarthy, M.},
  \bibinfo{year}{2007}.
\newblock \bibinfo{title}{Observation of relativistic electron precipitation
  during a rapid decrese of trapped relativistic electron flux}.
\newblock \bibinfo{journal}{Geophysical Research Letters} \bibinfo{volume}{34}.
\bibitem[{Millan and Thorne(2007)}]{Millan2007362}
\bibinfo{author}{Millan, R.}, \bibinfo{author}{Thorne, R.},
  \bibinfo{year}{2007}.
\newblock \bibinfo{title}{Review of radiation belt relativistic electron
  losses}.
\newblock \bibinfo{journal}{Journal of Atmospheric and Solar-Terrestrial
  Physics} \bibinfo{volume}{69}, \bibinfo{pages}{362--377}.
\bibitem[{Mironova et~al.(2015)Mironova, Aplin, Arnold, Bazilevskaya, Harrison,
  Krivolutsky, Nicoll, Rozanov, Turunen and Usoskin}]{Mironova2015}
\bibinfo{author}{Mironova, I.}, \bibinfo{author}{Aplin, K.},
  \bibinfo{author}{Arnold, F.}, \bibinfo{author}{Bazilevskaya, G.},
  \bibinfo{author}{Harrison, R.}, \bibinfo{author}{Krivolutsky, A.},
  \bibinfo{author}{Nicoll, K.}, \bibinfo{author}{Rozanov, E.},
  \bibinfo{author}{Turunen, E.}, \bibinfo{author}{Usoskin, I.},
  \bibinfo{year}{2015}.
\newblock \bibinfo{title}{Energetic particle influence on the earth’s
  atmosphere}.
\newblock \bibinfo{journal}{Space Science Reviews} , \bibinfo{pages}{96}.
Type = Article
\bibitem[{Mishev and Velinov(2014)}]{Mishev14d}
\bibinfo{author}{Mishev, A.}, \bibinfo{author}{Velinov, P.},
  \bibinfo{year}{2014}.
\newblock \bibinfo{title}{Influence of hadron and atmospheric models on
  computation of cosmic ray ionization in the atmosphere-extension to heavy
  nuclei}.
\newblock \bibinfo{journal}{Journal of Atmospheric and Solar-Terrestrial
  Physics} \bibinfo{volume}{120}, \bibinfo{pages}{111--120}.
\bibitem[{Neal et~al.(2015)Neal, Rodger, Clilverd, Thomson, Raita and
  Ulich}]{Neal20152194}
\bibinfo{author}{Neal, J.}, \bibinfo{author}{Rodger, C.},
  \bibinfo{author}{Clilverd, M.}, \bibinfo{author}{Thomson, N.},
  \bibinfo{author}{Raita, T.}, \bibinfo{author}{Ulich, T.},
  \bibinfo{year}{2015}.
\newblock \bibinfo{title}{Long-term determination of energetic electron
  precipitation into the atmosphere from aarddvark subionospheric vlf
  observations}.
\newblock \bibinfo{journal}{Journal of Geophysical Research A: Space Physics}
  \bibinfo{volume}{120}, \bibinfo{pages}{2194--2211}.
\bibitem[{O'Brien(1970)}]{Bri70}
\bibinfo{author}{O'Brien, K.}, \bibinfo{year}{1970}.
\newblock \bibinfo{title}{Calculated cosmic ray ionization in the lower
  atmosphere}.
\newblock \bibinfo{journal}{Journal of Geophysical Research}
  \bibinfo{volume}{75}, \bibinfo{pages}{4357--4359}.
\bibitem[{Park et~al.(2013)Park, Lee, Shin, Cho and Lee}]{Park2013247}
\bibinfo{author}{Park, M.Y.}, \bibinfo{author}{Lee, D.Y.},
  \bibinfo{author}{Shin, D.K.}, \bibinfo{author}{Cho, J.H.},
  \bibinfo{author}{Lee, E.H.}, \bibinfo{year}{2013}.
\newblock \bibinfo{title}{Dependence of energetic electron precipitation on the
  geomagnetic index kp and electron energy}.
\newblock \bibinfo{journal}{Journal of Astronomy and Space Science}
  \bibinfo{volume}{30}, \bibinfo{pages}{247--253}.
\bibitem[{Peck et~al.(2015)Peck, Randall, Green, Rodriguez and
  Rodger}]{Peck20154596}
\bibinfo{author}{Peck, E.}, \bibinfo{author}{Randall, C.},
  \bibinfo{author}{Green, J.}, \bibinfo{author}{Rodriguez, J.},
  \bibinfo{author}{Rodger, C.}, \bibinfo{year}{2015}.
\newblock \bibinfo{title}{Poes meped differential flux retrievals and electron
  channel contamination correction}.
\newblock \bibinfo{journal}{Journal of Geophysical Research A: Space Physics}
  \bibinfo{volume}{120}, \bibinfo{pages}{4596--4612}.
\bibitem[{Picone et~al.(2002)Picone, Hedin, Drob and Aikin}]{Picone2002}
\bibinfo{author}{Picone, J.}, \bibinfo{author}{Hedin, A.},
  \bibinfo{author}{Drob, D.}, \bibinfo{author}{Aikin, A.},
  \bibinfo{year}{2002}.
\newblock \bibinfo{title}{Nrlmsise-00 empirical model of the atmosphere:
  Statistical comparisons and scientific issues}.
\newblock \bibinfo{journal}{Journal of Geophysical Research: Space Physics}
  \bibinfo{volume}{107}, \bibinfo{pages}{1468}.
\bibitem[{Porter et~al.(1976)Porter, Jackman and Green}]{Porter1976154}
\bibinfo{author}{Porter, H.}, \bibinfo{author}{Jackman, C.},
  \bibinfo{author}{Green, A.}, \bibinfo{year}{1976}.
\newblock \bibinfo{title}{Efficiencies for production of atomic nitrogen and
  oxygen by relativistic proton impact in air}.
\newblock \bibinfo{journal}{The Journal of Chemical Physics}
  \bibinfo{volume}{65}, \bibinfo{pages}{154--167}.
\bibitem[{Roble and Ridley(1987)}]{Roble1987}
\bibinfo{author}{Roble, R.}, \bibinfo{author}{Ridley, E.},
  \bibinfo{year}{1987}.
\newblock \bibinfo{title}{An auroral model for the ncar thermospheric general
  circulation model (tgcm)}.
\newblock \bibinfo{journal}{Annales Geophysicae Series A-upper Atmosphere and
  Space Sciences} \bibinfo{volume}{5}, \bibinfo{pages}{369--382}.
\bibitem[{Rodger et~al.(2010)Rodger, Clilverd, Green and Lam}]{Rodger2010}
\bibinfo{author}{Rodger, C.}, \bibinfo{author}{Clilverd, M.},
  \bibinfo{author}{Green, J.}, \bibinfo{author}{Lam, M.}, \bibinfo{year}{2010}.
\newblock \bibinfo{title}{Use of poes sem-2 observations to examine radiation
  belt dynamics and energetic electron precipitation into the atmosphere}.
\newblock \bibinfo{journal}{Journal of Geophysical Research: Space Physics}
  \bibinfo{volume}{115}, \bibinfo{pages}{A04202}.
\bibitem[{Rodger et~al.(2007)Rodger, Clilverd, Thomson, Gamble, Sepp{\"a}l{\"a},
  Turunen, Meredith, Parrot, Sauvaud and Berthelier}]{Rodger2007}
\bibinfo{author}{Rodger, C.}, \bibinfo{author}{Clilverd, M.},
  \bibinfo{author}{Thomson, N.}, \bibinfo{author}{Gamble, R.},
  \bibinfo{author}{Sepp{\"a}l{\"a}, A.}, \bibinfo{author}{Turunen, E.},
  \bibinfo{author}{Meredith, N.}, \bibinfo{author}{Parrot, M.},
  \bibinfo{author}{Sauvaud, J.A.}, \bibinfo{author}{Berthelier, J.J.},
  \bibinfo{year}{2007}.
\newblock \bibinfo{title}{Radiation belt electron precipitation into the
  atmosphere: Recovery from a geomagnetic storm}.
\newblock \bibinfo{journal}{Journal of Geophysical Research: Space Physics}
  \bibinfo{volume}{112}, \bibinfo{pages}{A11307}.
\bibitem[{Rozanov et~al.(2005)Rozanov, Callis, Schlesinger, Yang, Andronova and
  Zubov}]{Rozanov20051}
\bibinfo{author}{Rozanov, E.}, \bibinfo{author}{Callis, L.},
  \bibinfo{author}{Schlesinger, M.}, \bibinfo{author}{Yang, F.},
  \bibinfo{author}{Andronova, N.}, \bibinfo{author}{Zubov, V.},
  \bibinfo{year}{2005}.
\newblock \bibinfo{title}{Atmospheric response to noy source due to energetic
  electron precipitation}.
\newblock \bibinfo{journal}{Geophysical Research Letters} \bibinfo{volume}{32},
  \bibinfo{pages}{1--4}.
\bibitem[{Schr{\o}ter et~al.(2006)Schr{\o}ter, Heber, Steinhilber and
  Kallenrode}]{Schroter2006}
\bibinfo{author}{Schr{\o}ter, J.}, \bibinfo{author}{Heber, B.},
  \bibinfo{author}{Steinhilber, F.}, \bibinfo{author}{Kallenrode, M.},
  \bibinfo{year}{2006}.
\newblock \bibinfo{title}{Energetic particles in the atmosphere: A monte-carlo
  simulation}.
\newblock \bibinfo{journal}{Advances in Space Research} \bibinfo{volume}{37},
  \bibinfo{pages}{1597--1601}.
\bibitem[{Sloan et~al.(2011)Sloan, Bazilevskaya, Makhmutov, Stozhkov,
  Svirzhevskaya and Svirzhevsky}]{Sloan201129}
\bibinfo{author}{Sloan, T.}, \bibinfo{author}{Bazilevskaya, G.},
  \bibinfo{author}{Makhmutov, V.}, \bibinfo{author}{Stozhkov, Y.},
  \bibinfo{author}{Svirzhevskaya, A.}, \bibinfo{author}{Svirzhevsky, N.},
  \bibinfo{year}{2011}.
\newblock \bibinfo{title}{Ionization in the atmosphere, comparison between
  measurements and simulations}.
\newblock \bibinfo{journal}{Astrophysics and Space Sciences Transactions}
  \bibinfo{volume}{7}, \bibinfo{pages}{29--33}.
\bibitem[{Solomon(1993)}]{Solomon1993}
\bibinfo{author}{Solomon, S.}, \bibinfo{year}{1993}.
\newblock \bibinfo{title}{Auroral electron transport using the monte carlo
  method}.
\newblock \bibinfo{journal}{Geophysical Research Letters} \bibinfo{volume}{20},
  \bibinfo{pages}{185--188}.
\bibitem[{Stozhkov et~al.(2009)Stozhkov, Svirzhevsky, Bazilevskaya, Kvashnin,
  Makhmutov and Svirzhevskaya}]{Stozhkov20091124}
\bibinfo{author}{Stozhkov, Y.}, \bibinfo{author}{Svirzhevsky, N.},
  \bibinfo{author}{Bazilevskaya, G.}, \bibinfo{author}{Kvashnin, A.},
  \bibinfo{author}{Makhmutov, V.}, \bibinfo{author}{Svirzhevskaya, A.},
  \bibinfo{year}{2009}.
\newblock \bibinfo{title}{Long-term (50 years) measurements of cosmic ray
  fluxes in the atmosphere}.
\newblock \bibinfo{journal}{Advances in Space Research} \bibinfo{volume}{44},
  \bibinfo{pages}{1124--1137}.
\bibitem[{Turunen et~al.(2009)Turunen, Verronen, Sepp{\"a}l{\"a}, Rodger, Clilverd,
  Tamminen, Enell and Ulich}]{Turunen2009}
\bibinfo{author}{Turunen, E.}, \bibinfo{author}{Verronen, P.},
  \bibinfo{author}{Sepp{\"a}l{\"a}, A.}, \bibinfo{author}{Rodger, C.},
  \bibinfo{author}{Clilverd, M.}, \bibinfo{author}{Tamminen, J.},
  \bibinfo{author}{Enell, C.F.}, \bibinfo{author}{Ulich, T.},
  \bibinfo{year}{2009}.
\newblock \bibinfo{title}{Impact of different energies of precipitating
  particles on nox generation in the middle and upper atmosphere during
  geomagnetic storms}.
\newblock \bibinfo{journal}{Journal of Atmospheric and Solar-Terrestrial
  Physics} \bibinfo{volume}{71}, \bibinfo{pages}{1176--1189}.
\bibitem[{Usoskin and Kovaltsov(2006)}]{Usoskin2006}
\bibinfo{author}{Usoskin, I.}, \bibinfo{author}{Kovaltsov, G.},
  \bibinfo{year}{2006}.
\newblock \bibinfo{title}{Cosmic ray induced ionization in the atmosphere: Full
  modeling and practical applications}.
\newblock \bibinfo{journal}{Journal of Geophysical Research: Atmospheres}
  \bibinfo{volume}{111}, \bibinfo{pages}{D21206}.
\bibitem[{Velinov et~al.(2013)Velinov, Asenovski, Kudela, Lastovi\v{c}ka,
  Mateev, Mishev and Tonev}]{Velinov2013}
\bibinfo{author}{Velinov, P.}, \bibinfo{author}{Asenovski, S.},
  \bibinfo{author}{Kudela, K.}, \bibinfo{author}{Lastovi\v{c}ka, J.},
  \bibinfo{author}{Mateev, L.}, \bibinfo{author}{Mishev, A.},
  \bibinfo{author}{Tonev, P.}, \bibinfo{year}{2013}.
\newblock \bibinfo{title}{Impact of cosmic rays and solar energetic particles
  on the earth's ionosphere and atmosphere}.
\newblock \bibinfo{journal}{Journal of Space Weather and Space Climate}
  \bibinfo{volume}{3}, \bibinfo{pages}{A14}.
\bibitem[{Velinov et~al.(2009)Velinov, Mishev and Mateev}]{Velinov2009}
\bibinfo{author}{Velinov, P.}, \bibinfo{author}{Mishev, A.},
  \bibinfo{author}{Mateev, L.}, \bibinfo{year}{2009}.
\newblock \bibinfo{title}{Model for induced ionization by galactic cosmic rays
  in the earth atmosphere and ionosphere}.
\newblock \bibinfo{journal}{Advances in Space Research} \bibinfo{volume}{44},
  \bibinfo{pages}{1002--1007}.
\bibitem[{Verronen et~al.(2011)Verronen, Rodger, Clilverd and
  Wang}]{Verronen2011}
\bibinfo{author}{Verronen, P.}, \bibinfo{author}{Rodger, C.},
  \bibinfo{author}{Clilverd, M.}, \bibinfo{author}{Wang, S.},
  \bibinfo{year}{2011}.
\newblock \bibinfo{title}{First evidence of mesospheric hydroxyl response to
  electron precipitation from the radiation belts}.
\newblock \bibinfo{journal}{Journal of Geophysical Research: Atmospheres}
  \bibinfo{volume}{116}, \bibinfo{pages}{D07307}.
\bibitem[{Whittaker et~al.(2013)Whittaker, Gamble, Rodger, Clilverd and
  Sauvaud}]{Whittaker20137611}
\bibinfo{author}{Whittaker, I.}, \bibinfo{author}{Gamble, R.},
  \bibinfo{author}{Rodger, C.}, \bibinfo{author}{Clilverd, M.},
  \bibinfo{author}{Sauvaud, J.A.}, \bibinfo{year}{2013}.
\newblock \bibinfo{title}{Determining the spectra of radiation belt electron
  losses: Fitting demeter electron flux observations for typical and storm
  times}.
\newblock \bibinfo{journal}{Journal of Geophysical Research: Space Physics}
  \bibinfo{volume}{118}, \bibinfo{pages}{7611--7623}.
\bibitem[{Wild et~al.(2010)Wild, Honary, Kavanagh and Senior}]{Wild2010}
\bibinfo{author}{Wild, P.}, \bibinfo{author}{Honary, F.},
  \bibinfo{author}{Kavanagh, A.}, \bibinfo{author}{Senior, A.},
  \bibinfo{year}{2010}.
\newblock \bibinfo{title}{Triangulating the height of cosmic noise absorption:
  A method for estimating the characteristic energy of precipitating
  electrons}.
\newblock \bibinfo{journal}{Journal of Geophysical Research: Space Physics}
  \bibinfo{volume}{115}, \bibinfo{pages}{A12326}.
\bibitem[{Wissing and Kallenrode(2009)}]{Wissing2009}
\bibinfo{author}{Wissing, J.}, \bibinfo{author}{Kallenrode, M.B.},
  \bibinfo{year}{2009}.
\newblock \bibinfo{title}{Atmospheric ionization module osnabr{\"u}ck (aimos): A
  3-d model to determine atmospheric ionization by energetic charged particles
  from different populations}.
\newblock \bibinfo{journal}{Journal of Geophysical Research: Space Physics}
  \bibinfo{volume}{114}, \bibinfo{pages}{A06104}.
\bibitem[{Wissing et~al.(2011)Wissing, Kallenrode, Kieser, Schmidt, Rietveld,
  Str{\o}mme and Erickson}]{Wissing2011}
\bibinfo{author}{Wissing, J.}, \bibinfo{author}{Kallenrode, M.B.},
  \bibinfo{author}{Kieser, J.}, \bibinfo{author}{Schmidt, H.},
  \bibinfo{author}{Rietveld, M.}, \bibinfo{author}{Str{\o}mme, A.},
  \bibinfo{author}{Erickson, P.}, \bibinfo{year}{2011}.
\newblock \bibinfo{title}{Atmospheric ionization module osnabr{\"u}ck (aimos): 3.
  comparison of electron density simulations by aimos-hammonia and incoherent
  scatter radar measurements}.
\newblock \bibinfo{journal}{Journal of Geophysical Research: Space Physics}
  \bibinfo{volume}{116}, \bibinfo{pages}{A08305}.

\end{thebibliography}

\end{document}